\documentclass{aa}
\usepackage{graphicx}
\usepackage{natbib}
\usepackage{amsmath}
\usepackage{txfonts}
\usepackage{ulem}

\begin{document}
   \title{Abell 611}
  
   \subtitle{I. Weak lensing analysis with LBC}

    \author{
     A. Romano\inst{1,2}, L. Fu\inst{3,12}, F. Giordano\inst{4,2}, R. Maoli\inst{1}, P. Martini\inst{5}, M. Radovich\inst{3,14}, R. Scaramella\inst{2},  V. Antonuccio-Delogu\inst{6,13}, A. Donnarumma\inst{7,8,9}, S. Ettori\inst{8,9}, K. Kuijken\inst{10,3}, M. Meneghetti\inst{8,9}, L. Moscardini\inst{7,9}, S. Paulin-Henriksson\inst{11,6}, E. Giallongo\inst{2}, R. Ragazzoni\inst{14}, A. Baruffolo\inst{14}, A. DiPaola\inst{2}, E. Diolaiti\inst{8}, J. Farinato\inst{14}, A. Fontana\inst{2}, S. Gallozzi\inst{2}, A. Grazian\inst{2}, J. Hill\inst{16}, F. Pedichini\inst{2}, R. Speziali\inst{2}, R. Smareglia\inst{15} \and V. Testa\inst{2}
   \institute{Dipartimento di Fisica, Universit\'{a} La Sapienza, piazzale A. Moro 2, I-00185 Roma, Italy\\
              \email{Anna.Romano@roma1.infn.it}
             \and
             INAF- Osservatorio Astronomico di Roma, via Frascati 33, 00044 Monte Porzio Catone (Roma), Italy 
	     \and		     
             INAF- Osservatorio Astronomico di Napoli, via Moiariello 16, 80131 Napoli, Italy 
             \and
	     Dipartimento di Fisica, Universit\'{a} Tor Vergata, Via della Ricerca Scientifica 1, I-00133 Roma, Italy
	     \and
	     Department of Astronomy and Center for Cosmology and Astroparticle Physics, The Ohio State University, Columbus, OH 43210, USA
	     \and
	     INAF- Osservatorio Astrofisico di Catania, Via Santa Sofia 78, I-95123 Catania, Italy	     
	     \and   
	     Dipartimento di Astronomia, Universit\'{a} di Bologna, via Ranzani 1, 40127 Bologna, Italy      
	     \and
	     INAF- Osservatorio Astronomico di Bologna, via Ranzani 1, 40127 Bologna, Italy 
	     \and
	     INFN, Sezione di Bologna, viale Berti Pichat 6/2, I-40127 Bologna, Italy
	     \and
	     Leiden Observatory, PO Box 9513, 2300 RA Leiden, The Netherlands     
	     \and
	     Service d'Astrophysique, CEA Saclay, Batiment 709, 91191 Gif-sur-Yvette Cedex, France
	     \and 
             Key Lab for Astrophysics, Shanghai Normal University, 100 Guilin Road, 200234, Shanghai, China
	     \and
	     Astrophysics, Department of Physics, University of Oxford, Oxford, UK
	     \and
	     INAF- Osservatorio Astronomico di Padova, vicolo dell'Osservatorio 5, I-35122 Padova, Italy 
	     \and
	     INAF- Osservatorio Astronomico di Trieste, Via G. B. Tiepolo 11, I-34131 Trieste, Italy
	     \and
	     Large Binocular Telescope Observatory, University of Arizona, 933 N Cherry Avenue, 85721-0065 Tucson, Arizona 
             }
	   }  

   \date{Received ..., 200?; ... 200?}

 
  \abstract
   {}
   {
   The Large Binocular Cameras (LBC) are two twin wide field cameras (FOV $\sim 23'\times25'$) mounted at the prime foci of the 8.4m Large Binocular Telescope (LBT). We performed a weak lensing analysis of the $z=0.288$ cluster Abell 611 on $g$-band data obtained by the blue-optimized Large Binocular Camera in order to estimate the cluster mass.
  }
   {Due to the complexity of the PSF of LBC, we decided to use two different approaches, KSB and Shapelets, to measure the shape of background galaxies and to derive the shear signal produced by the cluster. Then we estimated the cluster mass with both aperture densitometry and parametric model fits.}
   {The combination of the large aperture of the telescope and the wide field of view allowed us to map a region well beyond the expected virial radius of the cluster and to get a high surface density of background galaxies (23 galaxies/arcmin$^2$). This made possible to estimate an accurate mass for Abell 611. We find that the mass within 1.5 Mpc is: $(8\pm3)\times 10^{14} M_\odot$ from the
 aperture mass technique and $(5\pm1)\times 10^{14} M_\odot$ using the model fitting
 by a NFW mass density profile, for both Shapelets and KSB methods. This analysis demonstrates that LBC is a powerful instrument for weak gravitational lensing studies.}
   {}

   \keywords{Cosmology: dark matter; Galaxies: clusters: individual; Gravitational lensing}
   \titlerunning{Abell 611: I Weak lensing analysis with LBC}
   \authorrunning{A. Romano et al.}
   \maketitle
   
%

\section{Introduction}

\begin{figure}
\begin{center}
 \includegraphics[width=8cm]{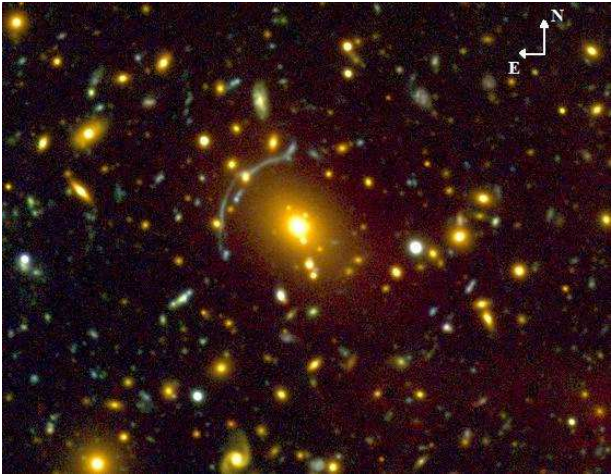}
 \caption{A three-color image ($2'\times1.6'$) of Abell 611 obtained with LBC observations in the $u$-, $g$-, and $r$-bands. 
 A giant arc is clearly visible close the brightest cluster galaxy  ($\alpha$= 08h 00m 57s,$\delta$=+36d 03' 23'').}\label{fig:ugr} 
\end{center}
\end{figure}

According to the hierarchical model of structure formation, clusters of galaxies 
are the most massive objects in the universe and the cluster mass function is a powerful probe of cosmological 
parameters (e.g. \citealt{evrard89,eke98,henry00,allen04,vikh09}). In addition, 
the ratio between the cluster gas mass, as estimated with X-ray observations, and the total mass in a galaxy  
cluster provides stringent constraints on the total matter density.   
Specifically, the apparent evolution of the gas fraction with redshift 
can be used to estimate the contribution of the dark energy component to 
the cosmic density (e.g \citealt{allen08,ettori09}).
The use of clusters as a cosmological probe therefore requires reliable mass estimates. \\ 
Several techniques are commonly used to estimate masses for galaxy
clusters: the X-ray luminosity or temperature of the hot intracluster
gas, the Sunyaev-Zel'dovich effect, the number of
bright galaxies in a cluster, and the velocity dispersion of the cluster galaxies.
The disadvantage of all of these methods is that they are indirect and require significant assumptions about the
dynamical state of the cluster. Gravitational lensing, in contrast, is
only sensitive to the amount of mass along the line of sight and allows 
reconstruction of the projected cluster mass regardless 
its composition or dynamical behavior (e.g. \citealt{KS93}). 
The only direct method to estimate cluster masses is therefore via measurement of the distortion (\textit{shear}) of the shapes of background galaxies that are weakly lensed by the gravitational potential of the cluster.\\
This distortion is very small and lensed galaxies are usually at high redshift. Observational studies to measure weak gravitational lensing by clusters require deep images in order to detect these faint sources and to obtain a high number density of background galaxies. Moreover this kind of analysis requires very high quality images in order to measure the shape of the lensed sources with high precision: good seeing ($ <1\arcsec$) conditions and a high signal-to-noise ratio (typically $>10$) are needed.  
Wide-field images are also required to obtain a statistical measure of the tangential
shear as function of distance from the cluster center so that the projected mass measured by weak lensing includes essentially all of the mass of the cluster (\citealt{clowe01,clowe02}).\\
In the last decade substantial progress has been made with weak lensing studies thanks to the
advent of wide-field data with linear detectors, the development of sophisticated algorithms for shape measurements 
(e.g. \citealt{KSB95}, \citealt{BJ02}, \citealt{refregier03}, \citealt{kui06}), and the availability of 
 multi-band photometry which provides information about the redshift distribution of the lensed sources \citep{Ilbert}. \\
Here we describe the results of a weak lensing analysis of the \object{Abell 611}
cluster. This analysis is based on images obtained with the Large Binocular Camera
(LBC), which are a pair of prime focus cameras mounted on the two 8.4m diameter mirrors of the Large Binocular Telescope (LBT). Each LBC has a $23'\times25'$ field of view (FOV) and, combined with the collecting area of LBT, is a 
very powerful instrument for weak lensing studies. \\
Abell 611 is a rich cluster at redshift $z=0.288$ \citep{red} that appears relaxed in X-ray data, has a regular morphology, 
and the brightest cluster galaxy (BCG) is 
coincident with the center of X-ray emission (Donnarumma et al. in preparation).
A giant arc due to strong lensing is also clearly visible close to the BCG (Fig. \ref{fig:ugr}). 
\\ 
In this work we describe a weak lensing analysis to estimate the 
mass of Abell 611 from a deep $g$-band LBC image whose field of view extends 
well beyond the expected virial radius of the cluster. 
We compare the mass estimated from gravitational lensing with previous lensing results
and with other mass estimates available in the literature that were derived by secondary techniques. 
In particular we compared our weak lensing results with new mass measurements
obtained by X-ray analysis of Chandra data provided to us by Donnarumma et al. (in preparation).\\
Mass measurements by weak lensing do not need any assumption about the geometry of the cluster;
however, assumptions are required to compare projected lensing masses with other mass estimates.  
Thus projection effects have to be taken into account during this kind of analysis, in particular 
the true triaxiality of the halo (\citealt{defilippis05}, \citealt{gavazzi05}) and the presence of unrelated structures 
along the line of sight (\citealt{metzler}, \citealt{hoekstra07})
can be sources of noise or bias on the projected mass measurements.\\
The paper is organized as follows. In the first sections we describe the data (Section \ref{data}) 
used for a weak lensing analysis of Abell 611, the catalog extraction of the background sources (Section~\ref{catalog}),
and the selection of candidate cluster galaxies (Section \ref{red}). The two different
pipelines used to extract the shear signal from the images are described in Section \ref{analysis}, and their 
results are compared in Section \ref{comparison}. Finally, both shear maps are used 
to estimate the mass of the cluster with different techniques (Section \ref{mass}).
The results are summarized and discussed in the Section \ref{summary}.\\
Throughout this paper we adopt $H_0$= 70 km $s^{-1}$ $Mpc^{-1}$, $\Omega_{m}$=0.3, and $\Omega_{\Lambda}$=0.7. At the distance of Abell 611, $11'$ radius corresponds to a projected physical distance of nearly 3 Mpc. 

\section{Observations and Data Reduction}\label{data}

\begin{figure}
\begin{center}
 \includegraphics[width=8cm]{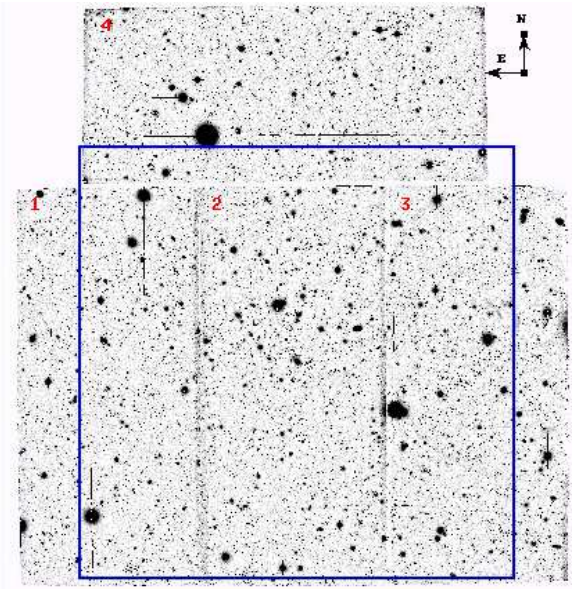}
 \caption{$g$-band image of the full field of LBC, centered on Abell 611. The box marks the region of 5000 pixels we used for the analysis.}\label{fig:box} 
\end{center}
\end{figure}

Abell 611 was observed in March 2007, during the Science Demonstration Time (SDT) for the blue-optimized Large Binocular Camera (LBC), which is one of the two LBCs built for the prime foci of the LBT. 
The LBC focal plane consists of four CCDs (2048 x 4608 pixels, pixel scale 13.5 $\mu$m, gain $\sim$ 2 e$^-$/ADU, readout-noise $\sim11$ e$^-$). The CCDs are arranged so that three of the chips are butted along their long edges and the fourth chip is rotated counterclockwise by 90 degrees and centered along the top of the other CCDs  (see Fig.\ref{fig:box}). The field of view is equivalent to $23'\times25'$ and provides images with a sampling of $0.225''$/pixel. Because each LBC is mounted on a swing arm over the primary mirror, the support structure lacks the symmetry of most prime focus instruments. Moreover LBC PSFs are dominated by optical aberrations from mis-alignments, which can cause PSFs to not have bilateral symmetry.
This is a potential complication for the weak lensing analysis and we discuss this point further below. More details about the characteristics of LBC are given in \citet{giallongo}.\\ 
The observations, collected in optimal seeing conditions (FWHM $\sim 0.6''$), consisted of several sets of exposures of 5 minutes each in a wide $u$-band, SDSS $g$- and $r$-band filters. 
The total exposure time was 1 hour in $g$, 15 minutes in $r$, and 30 minutes in $u$. Unfortunally some $u$-band observations were not usable, so the real total exposure time used for this band is 20 minutes. For the present work we used the deep, $g$-band data for the weak-lensing analysis and the $u$- and $r$-band data to select cluster galaxies. 
Each image was dithered by 5 arcseconds in order to remove bad pixels, rows, columns, and satellite tracks. This offset is not large enough to fill the gaps between the CCDs, but the analysis plan was to treat CCDs separately due to expected PSF discontinuities at the chip boundaries. The offsets were therefore kept small to maximize depth and uniformity. 
The images were reduced by the LBT pipeline\footnote{http://lbc.oa-roma.inaf.it} implemented at INAF-OAR. The flat-field correction was done using both a twilight flat-field and a superflat obtained during the night. Moreover a  geometric distortion correction was performed in order to normalize the pixel size, which showed differences across the CCDs due to field distortions in the optics. The astrometric solution was computed using the \textsc{ASTROMC} package \citep{rad08}. This solution was then used to resample and coadd the images using the \textsc{SWarp}\footnote{developed by E. Bertin, http://terapix.iap.fr} software. \\
Standard fields for photometric calibration were not observed during the SDT. 
 So we used the values of zero points for each band (Tab. \ref{tab:filters}) given by \citet{giallongo}.
Table \ref{tab:filters} also shows the limiting magnitudes estimated from the faintest point-like objects detected at the 5$\sigma$ and 20$\sigma$ level. 
\begin{table}
 \begin{center}
\begin{tabular}{cccccc}
\hline
Filter 	& $N_{exp}$ & exptime &     mag      &    mag       & zero point \\
        &           &   (s)   & ($5\sigma$)  & ($20\sigma$) &  \\
\hline \hline
g  & 12 & 3600 & 28.0 & 26.0 & 28.4 \\
r  &  3  &  900 & 26.0 & 24.2 & 27.7  \\ 
u  &  4  &  1200 & 27.2 & 25.0  &    27.1  \\
\hline
\end{tabular}
\end{center}
\caption{Exposure times, limiting magnitudes for point-like sources and zero points (AB) for the observations in each band.  \label{tab:filters} }
\end{table} 

\section{Catalog extraction}\label{catalog}

The detection of sources was performed using the \textsc{SExtractor} package \citep{bertin96}. Regions of
the image presenting potential problems, such as spikes and halos around bright
stars, were masked by visual inspection; sources inside such regions were discarded from the final catalog.
In addition, we removed sources located at the borders of each CCD, where the
SNR was lower due to the small dither offset. Finally, a very bright star
dominates one of the CCDs. We therefore decided to limit our analysis to a box (displayed in Fig.\ref{fig:box}) 
with a size of 5000 pixels (corresponding to $\sim18.7'$) centered on Abell 611.
Starting from this box of 350 arcmin$^2$, the effective area used for the analysis was 
 $\sim$ 290 arcmin$^2$ after removing all the masked regions (30\% due to
bright stars, 70\% due to regions between adjacent CCDs with no data or low S/N).\\
\begin{figure}
   \begin{center}
 \includegraphics[height=8cm,angle=-90]{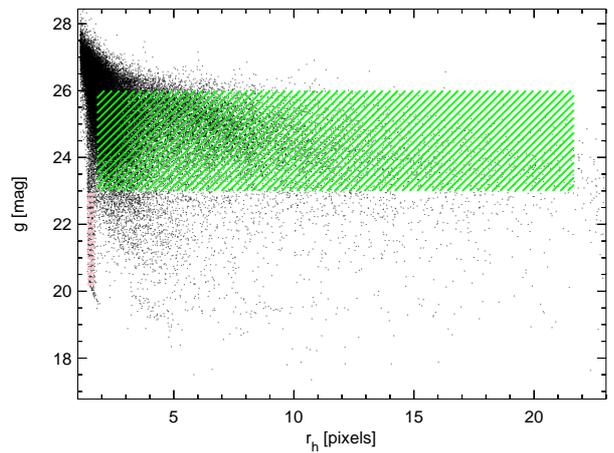}
\caption{Magnitude ($g$) vs. half-light radius ($r_h$) plane. \
\textit{Green zone} shows the background galaxies selected for
the lensing analysis ($23<g<26$, $r_h>1.8$ pixels) and \textit{pink
zone} the unsaturated stars selected for correction of PSF anisotropy
($20<g<23$, $1.4<r_h<1.8$ pixels). See \S\ref{catalog} for further details.}\label{fig.mag_rh}
\end{center}
\end{figure}
The separation between stars and galaxies was performed in the $mag-r_h$ plane,
where magnitudes (mag) and half-light radii ($r_h$) were obtained from the
MAG\_AUTO and FLUX\_RADIUS parameters computed by \textsc{SExtractor}. Unsaturated stars were
selected on the vertical branch (see Fig. \ref{fig.mag_rh}) in the range $20 < g < 23$ mag and
$1.4<r_h<1.8$ pixels. In this way we obtained 302 stars for the PSF
correction,  with a SNR$>200$ for the faintest ones.
\\
For the lensing analysis, only background galaxies located at redshifts larger
than $z=0.288$, the redshift of the cluster, should be used. Unfortunately, in our case  the number of available bands
does not allow us to estimate accurate photometric redshifts of these faint galaxies. 
The selection of the background galaxies was therefore done by choosing an
adequate cut in apparent magnitude.\\
The choice of the upper magnitude limit was based on the galaxy redshift
distribution obtained by \citet{Ilbert} from the Canada-France-Hawaii Telescope Legacy Survey
(CFHTLS), which also used the SDSS photometric system.
 Taking into account the accuracy of the photometric redshifts (3\%) of \cite{Ilbert}, the approximations
due to the different bands they used compared to ours and the assumption that we have the same galaxy distribution
in our field, we chose $0.4$ as redshift reference value to perform the magnitude cut.
Galaxies with $z \leq 0.4$ were assumed to belong to the cluster or be foreground galaxies.
 This reference value was chosen to be larger than cluster redshift, in order to reduce the contamination
of foreground galaxies as much as possible, taking into account the approximations discussed above.

\begin{figure}
\begin{center}
\includegraphics[width=8cm]{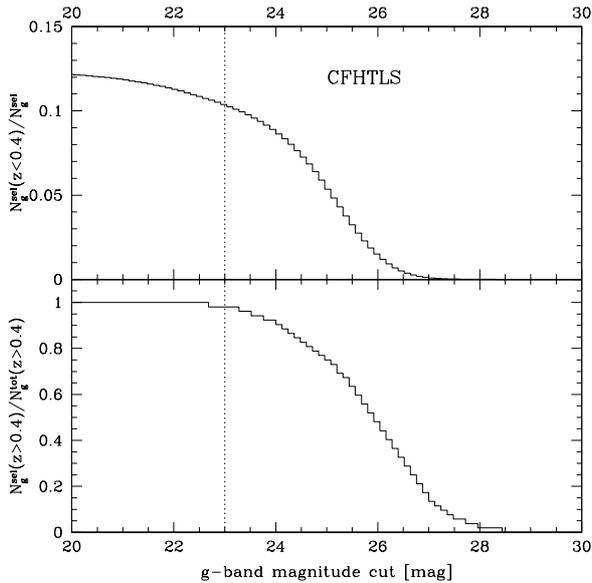}
\caption{The fraction of likely foreground and cluster galaxies (\textit{top-panel}) and background
galaxies (\textit{bottom-panel}) from the CFHTLS versus magnitude cut in $g$-band. The dotted vertical line 
is the lower limit of the magnitude cut adopted to select background galaxies. See \S\ref{catalog} 
for details.}\label{fig_CFHTLS}
\end{center}
\end{figure}

Figure \ref{fig_CFHTLS} shows the fraction of the total CFHTLS sources at
$z \leq 0.4$ ({\it upper panel}) and the fraction of background galaxies at $z > 0.4$ 
({\it bottom panel}) as a function of the apparent magnitude cut.
From this figure we conclude that a magnitude cut at $g>23$ is a good compromise to
minimize the contamination from likely foreground and cluster galaxies ($\sim10$\%) and to maximize
the number density of background galaxies ($\sim98$\%).  The faint magnitude cut was chosen at $g<26$, 
that is the magnitude limit where we have a signal-to-noise ratio SNR $>10$ for the sources, where 
SNR is defined as FLUX/FLUX\_ERR as measured by \textsc{SExtractor}.
The final catalog contained 8134 background galaxies. 

\section{Candidate cluster members}\label{red}
   
 The candidate cluster members were selected from the simultaneous usage of 
$u$-, $g$- and $r$- band photometry. To this end catalogs were extracted from these images, running SExtractor in dual-mode with the $g$-band image as detection image. Only sources detected in all three bands were used.

We then applied an algorithm (Fu et al., in preparation)
similar to the C4 Clustering Algorithm \citep{Miller05}.
 The algorithm is based on the assumption that galaxies in a cluster should have
 similar colors and locate together in space.  It evaluates the
 probability of each galaxy to be field-like: hence
candidate cluster galaxies are those for which this probability is below a given threshold, as outlined below.

\begin {enumerate}
\item Each galaxy was set in a four-dimensional space of $\alpha$, $\delta$,
  $u-g$ and $g-r$.  For each galaxy (named ``target
  galaxy''), we counted the number of neighbors within the
  four-dimensional box, $N_{\rm target}$. The angular size of the box
was set to 1 $h^{-1}$ Mpc ($\sim 5.5'$ at the redshift of Abell 611). 
The sizes of the boxes in two color dimensions were determined as
  \begin{equation}
    \delta_{mn} = \sqrt{\sigma_{mn}^2({\rm stat}) + \sigma_{mn}^2({\rm sys})},
  \end{equation}
  where $\sigma_{mn}({\rm stat})$ is the observed error for two
  magnitudes $(m,n)$, and $\sigma_{mn}({\rm sys})$ is the intrinsic
  scatter of the color $m-n$.  For LBC A611 data, $ \delta_{ug}$, $
  \delta_{gr}$ are 0.49 and 0.31 respectively.
\item This four-dimensional box was placed on 100 randomly chosen
  galaxy positions and at each position we counted the number of
  neighbors.  These randomized number counts constructed a
  distribution of counts for the target four-dimensional box. This
  distribution is represented by the median value $N_{\rm median}$ of
  the randomization counts.
\item The probability $p$ of the target galaxy being field-like was
  derived by comparing the target galaxy count $N_{\rm target}$ to the
  distribution of randomization values $N_{\rm median}$.

\item The distribution of $p$ values was derived by repeating the above
  steps for all galaxies. We ranked the $p$ values from smallest to
  highest and derived the value after which $p$ starts to rise
  significantly.  In this  way we identified $\sim$ 150 galaxies at $r < 23$ mag as 
  the candidate  cluster members. We further removed
  outliers in the $g-r$ vs $u-g$ diagram, leaving 125 candidate members.
  
\end {enumerate}

\begin{figure}
\centering
\includegraphics[width=7cm,angle=-90]{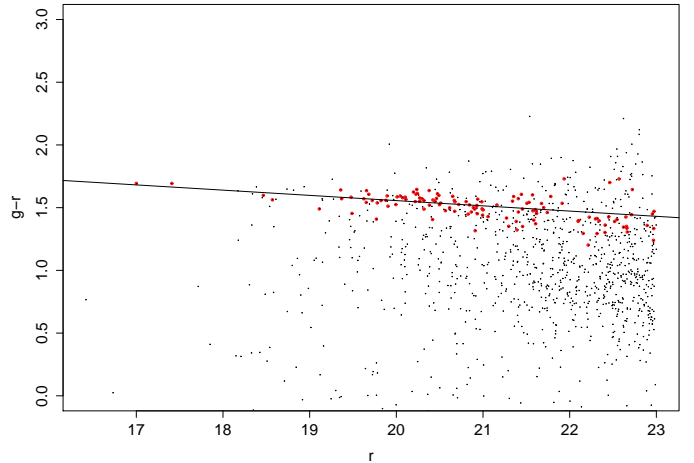}
\caption{Colour-magnitude plot of the galaxies in the Abell 611 field.
Red points are the candidate cluster galaxies
selected by the C4 method. Also displayed (solid line) is the results of
the biweight regression fit.}\label{fig:red_seq}
\end{figure}

Starting from these galaxies, we fitted the $g-r$ vs. $r$ Red Sequence ($g-r = a+b\cdot r$) using a biweight regression method (Fig. \ref{fig:red_seq}), and obtained: $a= 2.39_{-0.45}^{+0.61}$, $b=-0.04_{-0.03}^{+0.02}$. 

\section{Weak lensing analysis}\label{analysis}

Weak lensing is based on the measurement of the coherent distortion of
the shapes of background galaxies produced by a distribution of matter. This
distortion is very small and it is unmeasurable for any single background galaxy
because the galaxy intrinsic ellipticity is not known.  The dispersion of intrinsic ellipticity 
is a source of noise which is $\propto \sigma_\epsilon/\sqrt{N}$. A statistical approach is
therefore required, where the distortion can be measured for a large number of
sources  in order to bring down that noise. This requires a careful treatment of systematic effects, as the
shapes of galaxies may also be affected by contributions to the point spread function (PSF) by both the telescope and
the atmosphere. In the last decade several methods have been developed for this kind
of analysis. 
The most popular is the KSB approach, originally proposed by
\citet{KSB95} and improved by \citet{LK97} and \citet{hoekstra98}. Several different
implementations of this method exist in the literature and have been used in a number of KSB analysis pipelines. 
More recently, \citet{refregier03} and \citet{MR05} have proposed a 
new method based on Shapelets. 
Several available pipelines have also used this approach 
to measure the shear signal in various ways (e.g. \citealt{MR05}, \citealt{kui06}). \newline   
One of the key differences between these two approaches is the treatment of the PSF. 
The KSB method assumes that the PSF can be written as a convolution of a very
compact anisotropic kernel with a more extended, circular function. 
These two terms are expressed in terms of the quadrupole moments of the surface brightness. 
As summarized below, the isotropic component is subtracted from the measured ellipticity and 
the anisotropic component is subtracted from a responsivity term.
In contrast, in the Shapelets approach there is no assumption about the PSF shape. 
Individual galaxy images are decomposed into a complete orthonormal basis 
set consisting of Hermite (or Laguerre) Polynomials and the PSF correction is performed
through deconvolution. \\
Since the PSF of LBC presents a significant deviation from symmetry, the
analysis of Abell 611 presents a very good opportunity to compare the
results produced by these two methods. For this comparison we started with the same initial catalogs of stars and
galaxies for both pipelines. Specifically, we only considered sources with a
\textsc{SExtractor} FLAG $<$ 4, which removes sources that are possibly blended.
As the subsequent steps performed by each algorithm are different, the same 
galaxy may be rejected by one algorithm but not by the other. The result is 
that different output
catalogs are produced by the KSB and Shapelets pipelines. To have a homogeneous comparison,  we
therefore finally  selected only the sources common to both output catalogs.
Sources with an unphysical ellipticity $|e|>1$ were also removed from the final 
catalogs.

\subsection{KSB method}\label{ksb}

We used the weak lensing pipeline described in \citet{rad08} to compute the quantities relevant to the lensing analysis. This pipeline implements 
the KSB approach using a modified version of the IMCAT\footnote{http://www.ifa.hawai.edu/$\sim$kaiser/imcat}
tools that was provided to us by T. Erben (\citealt{erben01} and \citealt{hetter07}). \newline
In the KSB approach stars and galaxies are parametrized according to the 
weighted quadrupole moments of the intensity distribution using a Gaussian weight
function whose scale length is the size of the source  (the formalism is described in \citealt{KSB95}).
The main assumption of this approach is that the PSF can be described as the sum
of a large isotropic component (seeing) and small anisotropic part. 
In this way the observed ellipticity $e_{obs}$ can be related to the intrinsic
source ellipticity $e_{s}$ and shear $\gamma$ by the relation:
 \begin{equation}
  e_{obs}=e_s+P^{\gamma}\gamma+P^{sm}p,
 \end{equation}
where $P^{sm}$ is the smear polarizability tensor and $P^{\gamma}$ is the 
pre-seeing polarizability tensor, which is related to the shear polarizability 
tensor $P^{sh}$ and $P^{sm}$ by Eq.~\ref{pgamma} \citep[see also][]{hoekstra98}. 
The quantity $p$ characterizes the anisotropy of the PSF and is estimated from
stars, which have zero intrinsic ellipticity:
\begin{equation}\label{p}
\centering
  p^{*}=\frac{e_{obs}^{*}}{P^{sm*}}.
\end{equation}
If we average over a large number of sources, assuming a random orientation of the
unlensed galaxies, we expect $\langle e_s \rangle$ = 0 and so
\begin{equation}
 \gamma \ = \ \langle\frac{e_{iso}}{P^{\gamma}}\rangle,
\end{equation}
where $e_{iso}=e_{obs}-P^{sm}p^{*}$ is the 
ellipticity corrected for anisotropic distortions.\\

\begin{figure*}
   \begin{center}
 \includegraphics[angle=-90,width=0.9\textwidth]{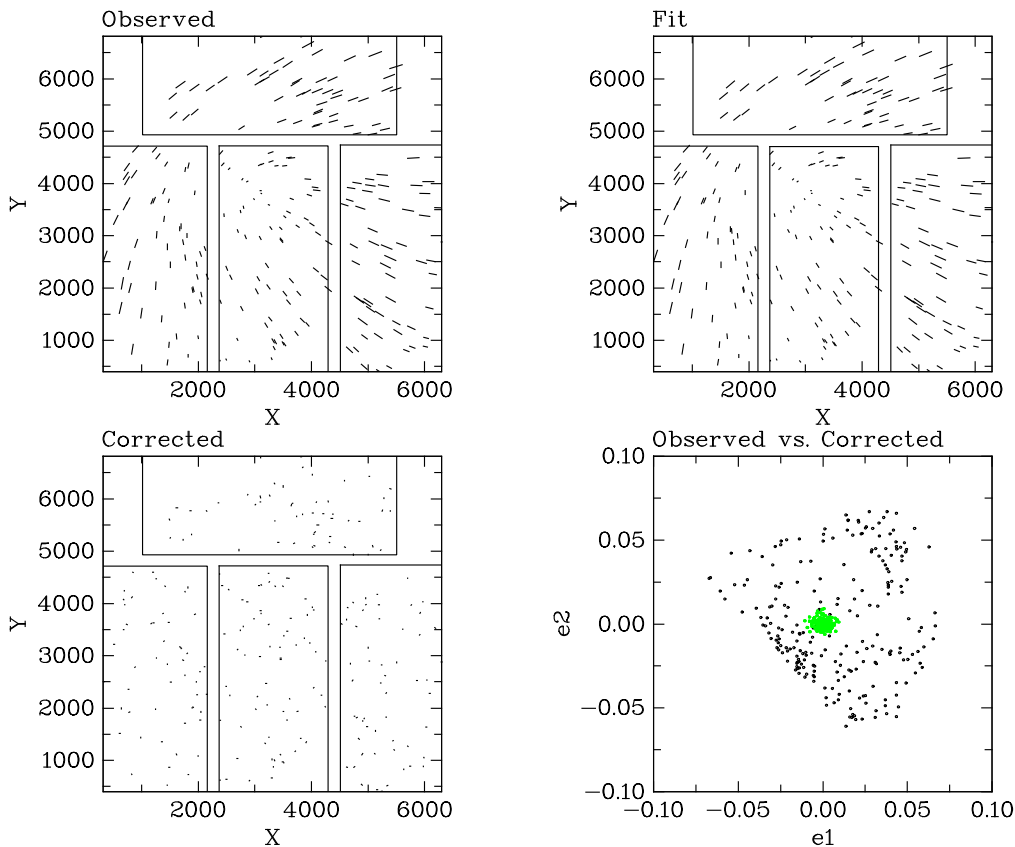}
\caption{Removal of PSF anisotropy in the KSB approach. The observed (\textit{top-left}), fitted (\textit{top-right}) and
residual (\textit{bottom-left}) ellipticities of stars for all CCDs. The 
observed (black points) versus corrected (green points) ellipticities are also shown (\textit{bottom-right}).}\label{fig_psf}
\end{center}
\end{figure*}

We computed the contribution due to the PSF anisotropy (Eq. \ref{p}) from the
selected stars. This quantity changes with position in the image, so we
needed to fit it on each CCD in order to extrapolate its value at the position
of the galaxy we want to correct. In our case we  performed this fitting on each CCD and 
a second-order polynomial fit was sufficient.\\ 
Fig. \ref{fig_psf} displays the spatial pattern of the ellipticities 
for the stars in all CCDs, before and after the PSF correction. \\
After that we computed $P^{\gamma}$ for each source: 
\begin{equation}\label{pgamma}
 P^{\gamma}= P^{sh}-P^{sm}\frac{P^{sh*}}{P^{sm*}}.
\end{equation}
As the PSF correction was done separately on each CCD, we decided not to fit $\frac{P^{sh*}}{P^{sm*}}$ as a function of the coordinates and instead we took an average value.
\begin{figure}
 \begin{center}
 \includegraphics[width=7cm,angle=270]{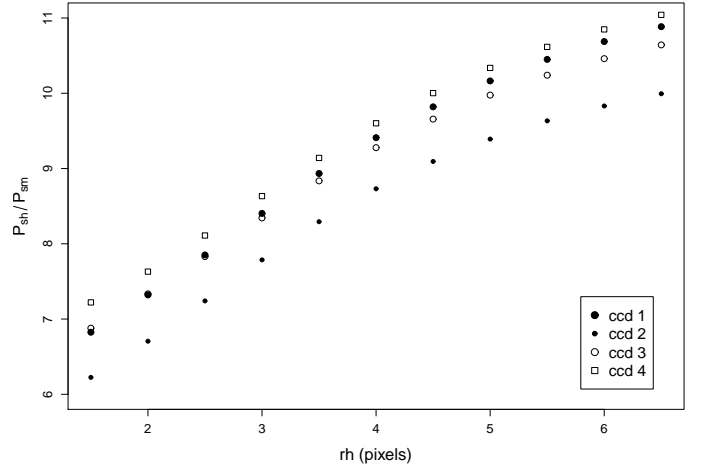}
 \caption{${P^{sh*}}/{P^{sm*}}$ values computed in different bins of $r_h$, for each CCD. These values are lower for the central CCD (\#2) where the quality of PSF is better.}\label{fig:pshsm}
\end{center}
\end{figure}

Stellar ellipticities should be computed using the same weight function used for galaxies (see \citealt{hoekstra98}), so we considered a sequence of bins in $r_{h}$ and for each galaxy we selected the PSF correction terms computed in the closest $r_{h}$ bin. Fig. \ref{fig:pshsm} shows the $\frac{P^{sh*}}{P^{sm*}}$ values computed in different bins of $r_h$, for each CCD. \\
We weighted the shear contribution from each galaxy according to:
\begin{equation}\label{w}
 w=\frac{P^\gamma{}^2}{P^\gamma{}^2 \sigma_{e0}{}^2 + \langle \Delta e^{2} \rangle},
\end{equation}
as in \citet{hoekstra00}, where $\sigma_{e0} \sim 0.3$ is the intrinsic rms of galaxy ellipticities, 
$\langle\Delta e^{2}\rangle^{1/2}$ is the uncertainty in the measured ellipticity, where the average is computed 
on both components of ellipticity for each galaxy (see eq. A8, A9 in \citet{hoekstra00}). \\
A crucial point in this kind of study is the selection of the galaxies to use
for the shear analysis, as the contamination of foreground galaxies can dilute
the lensing signal and lead to an underestimate of the mass.\\
Since the PSF degrades somewhat at the borders of the image, we limited our
analysis to an $18.7' \times 18.7'$ box centered around the BCG, 
as discussed in Section \ref{catalog}. 
After that we filtered the source catalog using the following criteria:
$P^{\gamma}>0.25$,  SNR $>10$, $r_h >$ 1.8 pixels, $23 < g < 26$, ellipticities
smaller than one, obtaining a surface density of $\sim$ 25 galaxies/arcmin$^2$. \\
The cut $P^{\gamma}>0.25$ allowed us to discard sources that appeared too 
circular (e.g.
stars incorrectly classified as galaxies). 
We considered only galaxies with SNR $> 10$ in order to avoid noisy
objects, which can be a source of error in the computation of the shear signal.
Finally, we used the magnitude cut $23 < g < 26$ to select background 
galaxies, whose choice was previously explained in Section \ref{catalog}.

\subsection{Shapelets}\label{KK}
\begin{figure*}
\begin{center}
 \includegraphics[height=0.7\textwidth, angle=0]{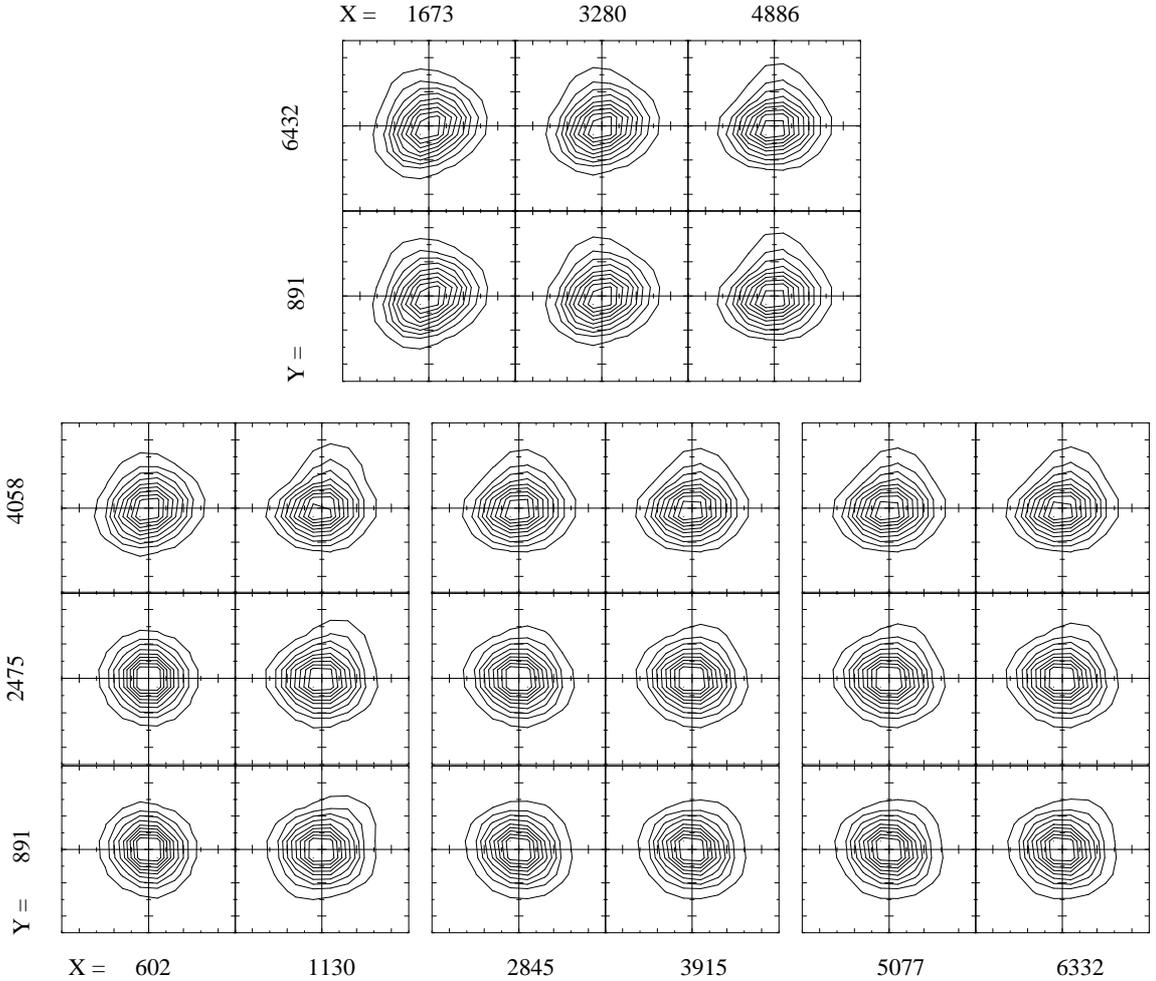}
 \caption{ Shapelets PSF models interpolated at different positions on
   each CCD. The distribution of these models corresponds to their
   actual placement on the CCD mosaic,    CCD1 to CCD3 from left to
     right in the bottom panel and CCD4 in the top. Contours are the
     representations of PSF shape decomposed by Shapelets. The X and Y
     values correspond to the pixel position.}\label{fig:psf_shape}
\end{center}
\end{figure*}

\begin{figure*}
\begin{center}
 \includegraphics[height=0.6\textwidth, angle=270]{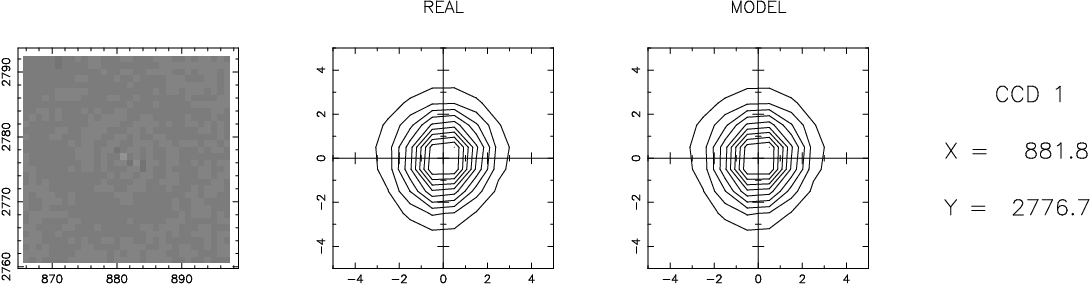}
 \includegraphics[height=0.6\textwidth, angle=270]{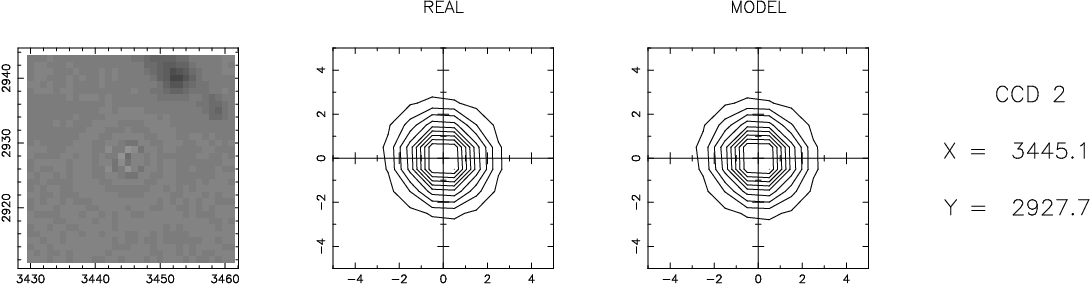}
 \includegraphics[height=0.6\textwidth, angle=270]{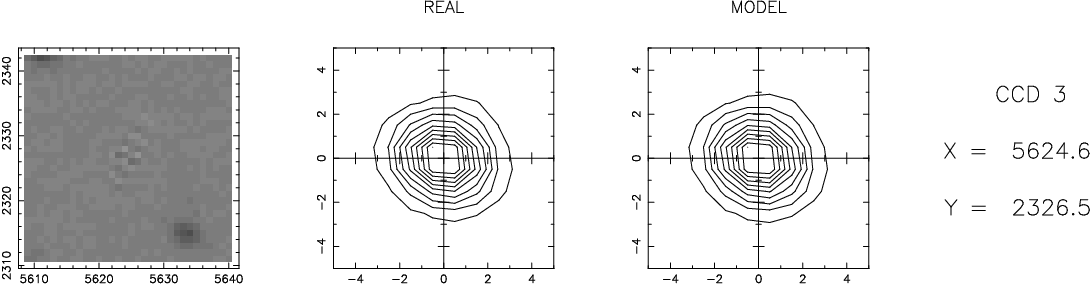}
 \includegraphics[height=0.6\textwidth, angle=270]{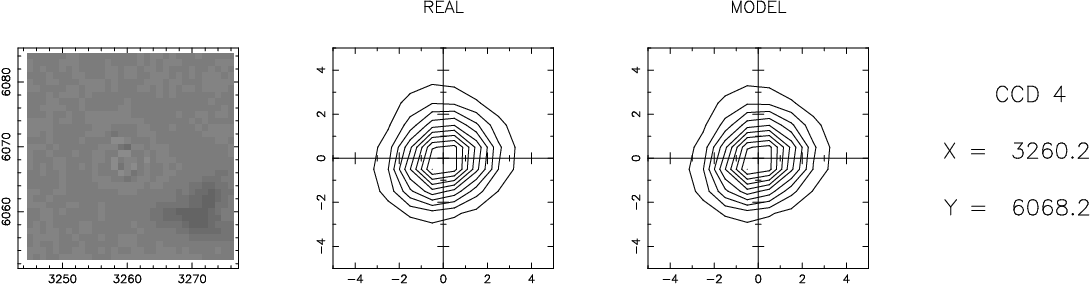}
\caption{ The star at the position near the center of each CCD are taken
  as an example to show the residual of PSF model fitting (left
    column). The real PSF shape and fitted PSF model are shown in the
   middle and right columns.  }\label{fig:residual}
\end{center}
\end{figure*}

Another approach for weak lensing analysis is the use of Shapelets,
which are basis functions constructed from two-dimensional
Hermite polynomials weighted by a Gaussian. The translation,
magnification, rotation and shear of astronomical images can be
expressed as matrices acting on Shapelets coefficients. The advantage
of Shapelets is that a galaxy image can be described in reverse order:
pixelation, convolution with the PSF, and finally distortion by shear. Shapelets
have a free scale radius $\beta$ which is the size of the Gaussian
core of the functions. Its truncated expansion describes deviations
from a Gaussian over a particular range of spatial scales, which
widens with order $N$. \\
For the analysis of the LBC data of Abell 611 we used the Shapelets pipeline
developed by \citet{kui06}, starting with the 
same stars and galaxies used in the KSB method. 
Each star was first fit with a
circular Gaussian and the median radius was computed. This radius 
was then multiplied by a factor of 1.3 for the Shapelets fits, which 
was found by \citet{kui06} to work well for a range of model PSFs up to Shapelets order $N=8$. 
Then we obtained a Shapelets description of the PSF for each star. In
order to estimate the PSF model at the position of each galaxy, the
Shapelets coefficients were interpolated by a fourth-order polynomial
on the whole image frame (Fig. \ref{fig:psf_shape}).    The residual of
the PSF model fitting is shown in Fig.~\ref{fig:residual}.  \\ The ellipticity
of each source is then determined by    least-squares fitting a
  model, which is expressed as the shear applied to a circular source
  to fit the object optimally.
   The extension order of Shapelets for galaxies is taken the same
  as for stars.  Performing the least-squares fit, the minimum of
  $\chi^2$ can be found in a few Levenberg-Marquardt iterations
  \citep{Press86}. The errors of Shapelets coefficients for each source
  are derived from the photon noise, and these can be propagated
  through in the $\chi^2$ function. Thus the error of shear
  measurement $\sigma_\gamma$ is calculated from the covariance matrix
  which is given by the second
  partial derivatives of $\chi^2$ at the best fit.   \\
The shear contribution from each galaxy is weighted according to:
\begin{equation}
w=\frac{ \sigma_{e0}^2}{\sigma_{e0}^2+\sigma_\gamma^2},
\end{equation} 
which combines the error in the shear measurements $\sigma_\gamma$ and the intrinsic scatter $\sigma_{e0}$. 
   The last quantity was computed from the distribution of ellipticity components of galaxies.\\  
Finally, we removed galaxies with 
SNR $<10$ and that failed the Shapelets expansion and radial profile cuts 
defined by \citet{kui06}. This eliminated galaxies that were not well detected or measured,
providing a number density of $\sim 26$ galaxies/arcmin$^2$. 
Further details about the selection criteria are available in \citet{kui06}. 

\subsection{Comparison between KSB and Shapelets ellipticities}\label{comparison}

After matching the two output catalogs, our final background galaxy sample 
has a surface density of $\sim$ 23 galaxies/arcmin$^2$.\\
   In weak lensing studies, the background number density of
  ground-based telescope usually is around 15 to 20 galaxies/arcmin$^2$ (e.g. \citealt{PH07}). It
  depends not only on the size of telescope, the exposure time, the
  color filter, seeing condition, but also on galaxy selection
  criteria.\\ 
The maps of the PSF correction computed by both methods show a good-quality 
PSF in the central regions that then degrades further from the center of the 
field. Nevertheless the final correction is
good with fluctuations in the PSF anisotropy less than 1\%.

\begin{figure}
 \begin{center}
 \includegraphics[width=7cm]{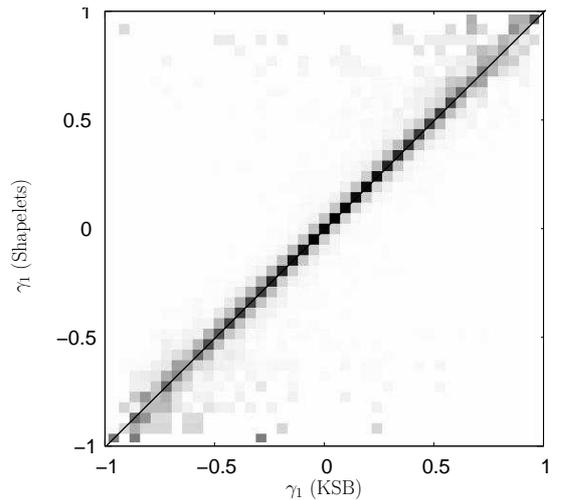}
\caption{Comparison between the first component of the shear from the Shapelets
  and KSB methods (results with the second components are very
  similar).}\label{fig:g}
 \end{center} 
\end{figure}
 
\begin{figure}
 \begin{center}
 \includegraphics[width=10cm]{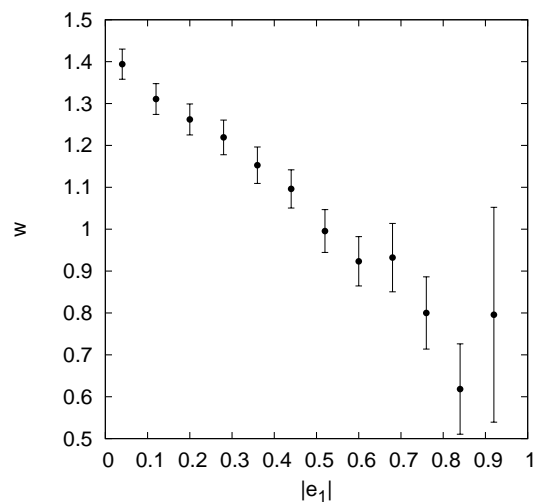}
\caption{The plot of the normalized galaxy weight as function of the absolute first component of ellipticity for KSB measurement. 
The applied normalization was $\Sigma w_i \times \delta e_1=1$, where $\delta e_1$ is
the constant bin width of $e_1$. The Shapelets measurement shows similar down-weight for elongated ellipticity galaxies.}\label{fig:weight}
 \end{center} 
\end{figure}

Figure \ref{fig:g} compares the first component of the shear
$\gamma_1$ measured by the KSB and Shapelets methods. It shows good
agreement in the range of $-0.5 < \gamma_1 < 0.5$. Some scatter is present 
for very elongated galaxies, but the fraction of these galaxies is
less than 5\% in the final, common catalog. As shown in
   Fig.~\ref{fig:weight}, these strongly elongated
galaxies are down-weighted nearly by a factor of 2 compared to small ellipticity
galaxies, so that they do not affect the final mass measurements of
the cluster (as described further below).  A similar behavior is seen
for the second component $\gamma_2$.  These matched catalogs were used
to compute estimates of the cluster mass as described in the next
sections.

\section{Mass Measurements}\label{mass}

The relationship between the shear $\gamma$ and the surface mass
density is:
\begin{equation}
 \gamma(\theta)=\frac{1}{\pi}\int_{R^2}d^2\theta^{\prime}D(\theta-\theta^{\prime})\kappa(\theta^{\prime}),
\end{equation}
where: $D=\frac{-1}{(\theta_1-i\theta_2)^2}$, $\kappa\equiv\Sigma/\Sigma_{cr}$ is the convergence, $\Sigma_{cr}=\frac{c^2}{4\pi G}\frac{D_s}{D_lD_{ls}}=\frac{c^2}{4\pi G}\frac{1}{D_l\beta}$ 
is the critical density of the cluster, and $D_s$, $D_l$
and $D_{ls}$ are source-observer, lens-observer and lens-source distances
respectively. While weak lensing actually measures the reduced shear $g=\frac{\gamma}{1-\kappa}$, for $\kappa\ll1$, $g\sim\gamma$.\\ 
An optimal approach for computing the mass requires knowledge of the redshift of each
background galaxy. As we do not have this information, we assumed that the
background sources all lie at the same redshift according to the
\textit{single sheet approximation} \citep{KS01}. An estimate of the redshift
value to use for the weak lensing analysis was computed using the first release
of photometric redshifts available for the D1 deep field of the CFHTLS,
adopting the magnitude cut in the $g$-band chosen here for the background galaxies selection ($23 < g < 26$ mag; see Section~\ref{catalog}) and
assuming a Gamma probability distribution \citep{gavazzi04}. This yielded a 
median redshift $z=1.05$ and for our analysis we assume all of the background 
galaxies lie at $z\sim1$, corresponding to 
$\langle\beta\rangle\ = \langle D_{ls}/D_s \rangle\ \sim \ 0.65$.\\
As discussed by \citet{hoekstra07}, the single-sheet approximation results in an overestimate of
the shear by a factor
\begin{equation}
1+\left[\frac{\langle\beta^2\rangle}{\langle\beta\rangle^2}-1\right] \kappa.
 \end{equation}
We computed this factor using the CFHTLS catalog of photometric redshifts and obtained $\langle\beta^2\rangle/\langle\beta\rangle^2 = 1.167$.\\
\\
The convergence $\kappa$ gives an estimate of the surface mass density apart from an
unknown additive constant -- the so-called \textit{mass-sheet degeneracy}. We
tried to solve this degeneracy using two different approaches: assuming either
that $\kappa$ vanishes at the borders of the image or a particular mass profile
whose expected shear profile is known.

\subsection{S-Maps}\label{s-map}

\begin{figure*}
 \begin{center}
 \includegraphics[viewport=60 20 470 400,width=7cm,clip]{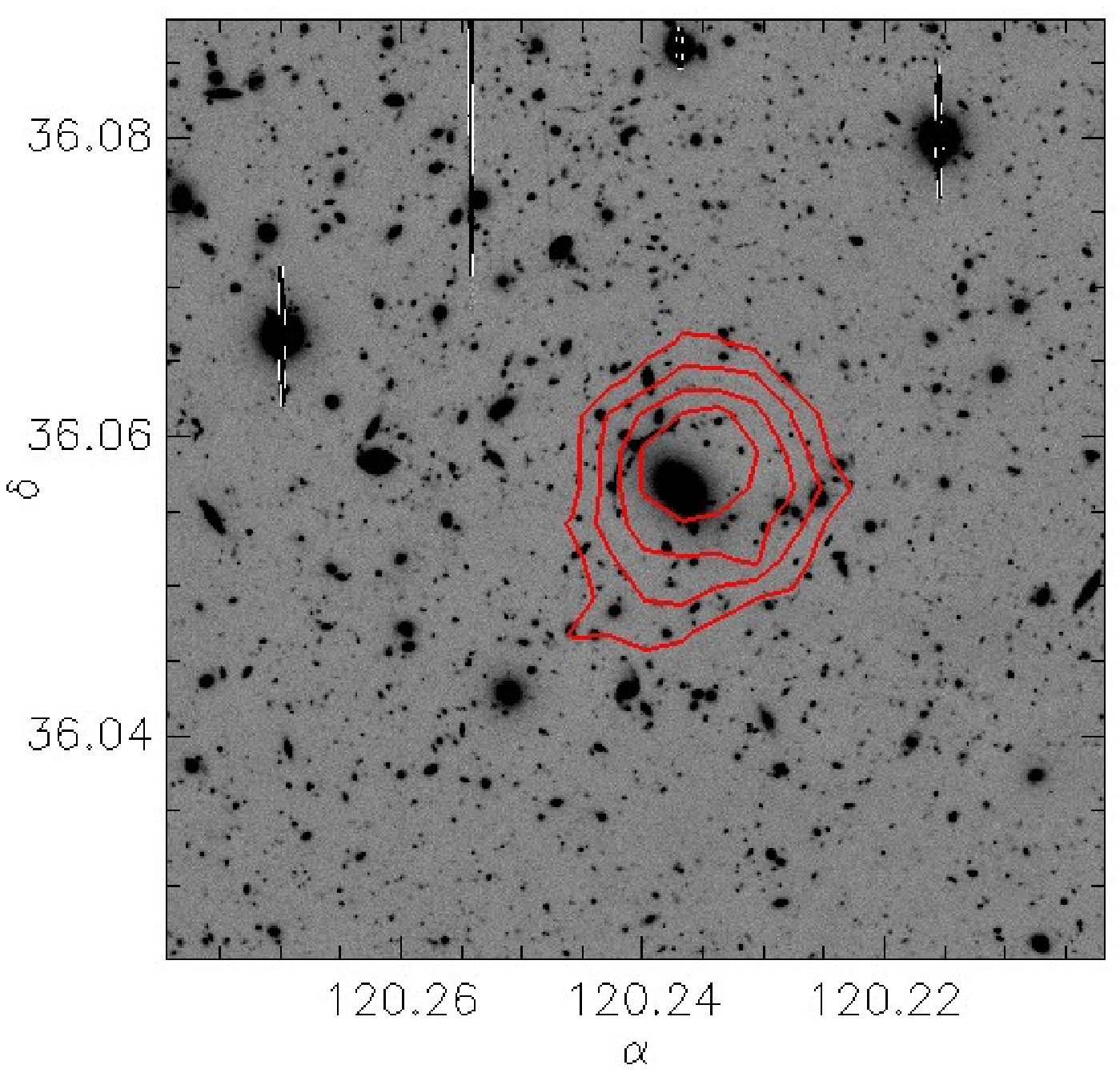}\includegraphics[viewport=60 20 470 400,width=7cm,clip]{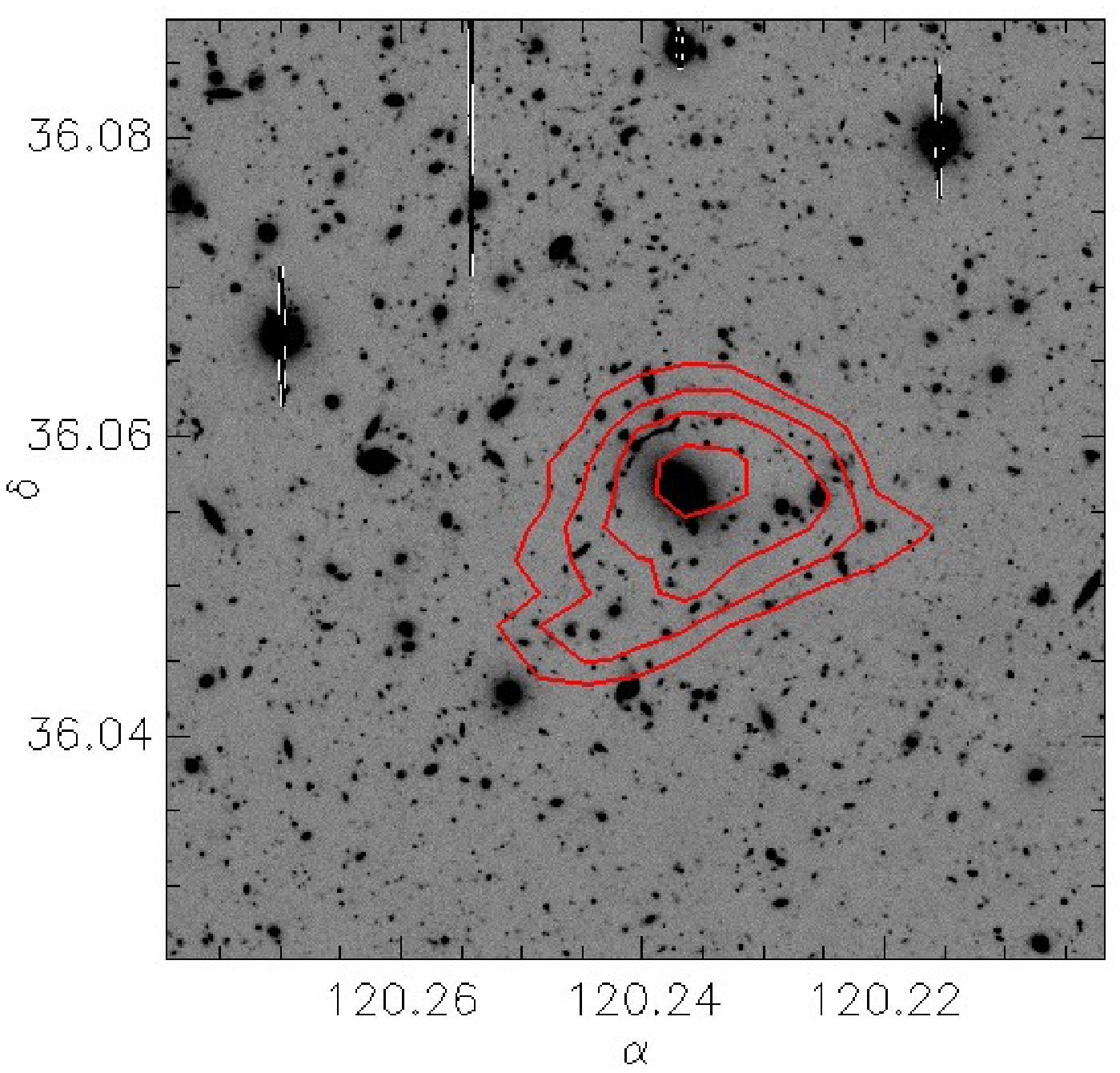}\\
\caption{S-maps obtained from the KSB (\textit{left panel}) and the Shapelets (\textit{right panel}) analysis. The levels are
plotted between $\sigma_{\rm min}=3.5$ and $\sigma_{\rm max}=5$. They are overplotted 
on a $g$-band greyscale image ($\sim 4$ arcmin) of the center of the field of Abell 611. }\label{fig:snr_ksb}
 \end{center}
 \end{figure*}

In Figure \ref{fig:snr_ksb} we plot the so-called
\textit{S-maps} \citep{schirm04} for these data. \textit{S-maps} are computed 
as the ratio $S=M_{ap}/\sigma_{Map}$ where: 
\begin{eqnarray}
 M_{ap}&=&\frac{\Sigma_ie_{t,i}w_iQ(|\theta_i-\theta_0|)}{\Sigma_iw_i}\\
 \sigma_{M_{ap}}^2&=&\frac{\Sigma_ie_{t,i}^2w^2_iQ^2(|\theta_i-\theta_0|)}{
2(\Sigma_iw_i)^2},
\end{eqnarray}
For this calculation the image is considered as a grid of points, $e_{t,i}$ 
are the tangential components of the ellipticities of the lensed galaxies, which are 
computed by considering the center of each point of the grid, $w_i$ is the weight as 
defined in Equation \ref{w}, and $Q$ is a window function, chosen to be a Gaussian function  defined by:
\begin{equation}
 Q(|\theta-\theta_0|)= \dfrac{1}{\pi \theta_c^2}exp\left( -\dfrac{(\theta-\theta_0)^2}{\theta_c^2}\right) 
\end{equation}
 where $\theta_0$ and $\theta_c$ are the center and the size of the aperture.
The ratio $S=M_{ap}/\sigma_{M_{ap}}$ provides an estimate of the SNR ratio of
the dark matter halo detection.
\textit{S-maps} are discussed further by \citet{schirm04}.

\begin{figure}
\centering
\includegraphics[width=7cm,angle=-90]{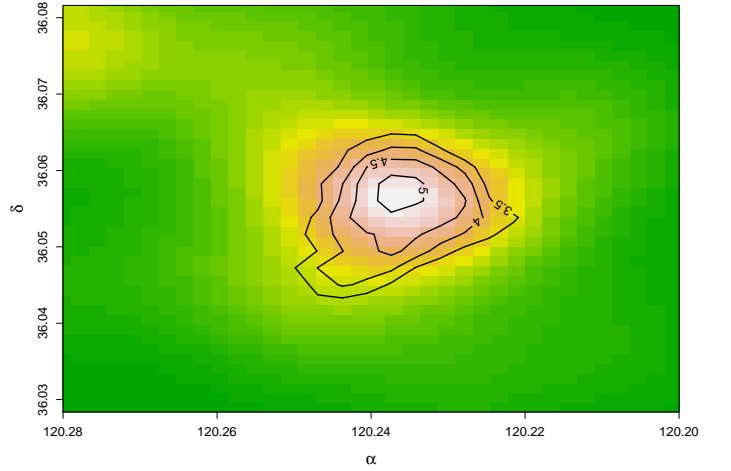}
\caption{$r$-band luminosity-weighted density distribution of red sequence galaxies of Abell 611. The overlaid contour (black lines) is the S-map (computed and discussed in Section \ref{s-map}) showing the SNR of the shear signal around the
cluster obtained from the Shapelets analysis. The levels are plotted between $\sigma_{\rm min}=3.5$ and $\sigma_{\rm max}=5$.}\label{fig:red_map}
\end{figure}

We computed these maps using shear catalogs obtained from both the KSB and Shapelets
pipelines. In both cases (see Fig. \ref{fig:snr_ksb}) 
the maps show that the lensing signal is peaked around the BCG, confirming 
that this is indeed the center of the mass distribution. The mass distribution also appears quite 
regular, which is in agreement with what is indicated by the X-ray maps. 
In Figure \ref{fig:red_map} S-map contours are overlaid on the $r$-band luminosity-weighted density distribution of the red sequence galaxies of Abell 611, selected in Section \ref{red}, showing that the mass distribution follows that of the red cluster galaxies.

\subsection{Aperture densitometry}

In order to trace the mass profile of the cluster, we computed the
$\zeta$-statistic described in \citet{clowe98} and \citet{fahlman94}: 
\begin{eqnarray}
\zeta(\theta_1)=\bar{\kappa}(\theta\le\theta_1)&-&\bar{\kappa}(\theta_2<\theta\le\theta_{max})=2\int^{\theta_2}_{\theta_1}\langle\gamma_T\rangle d\ln{\theta}\\ \nonumber
&+&\frac{2}{1-(\theta_2/\theta_{max})^2}\int^{\theta_{max}}_{\theta_2}\langle\gamma_T\rangle d\ln{\theta}, 
\end{eqnarray}\label{apdens}
Mass measurements are computed within different apertures of increasing radius 
using a control annulus far from the center of the distribution (the BCG).
$M_{ap}=\pi\theta_1^2\zeta(\theta_1)\Sigma_{cr}$ is the mass within the last
aperture and provides a lower limit on the mass, unless the value in the control
annulus is equal to zero. 
This method allows us to choose the size of the annulus that satisfies the
desired condition; moreover it has the advantage that this mass computation does 
not depend on the mass profile of the cluster
\citep{clowe98}. \\
We chose $30''\leq \theta \leq 500''$ and $\theta_{max}=600''$, which yielded 
projected masses within $\sim 1500$ kpc of $7.7\pm{3.3}\times 10^{14} M_{\odot}$ and $8.4\pm{3.8}\times 10^{14} M_{\odot}$ using the KSB and Shapelets shear catalogs, respectively.\\
Unfortunately we could not extend our analysis further from the center of the cluster because of the 
presence of a very bright star in the field that made the outer regions unusable.\\
Since weak lensing is sensitive to the total mass along the line of sight, the observed 
aperture mass is the sum of the mass of the cluster and any contribution
from other, uncorrelated structures along the line of sight. 
This contribution is assumed to be negligible in the central regions of the cluster, which are much denser,
and become more relevant in its outer regions. As discussed by \cite{hoekstra01}, the effect of this
contribution does not introduce any bias, but it does add a source of noise to the lensing mass. 
Aperture densitometry is more affected by this uncertainty than parametric methods because it is sensitive to the lensing signal at large radii.
Nevertheless, for observations of rich clusters at intermediate redshifts, this uncertainty is fairly small because the bulk of the background sources are at much higher redshifts than the cluster.

\subsection{Model fitting}\label{fit}

The model fitting approach for estimating the mass of a cluster consists of 
assuming a particular analytic mass density profile for which calculate the 
expected shear and then fitting the observed shear with the model 
by minimizing the log-likelihood function (\citealt{schneider00}): 
\begin{equation}
l_{\gamma}=\sum_{i=1}^{N_{\gamma}}\left[\frac{|\epsilon_i-g(\theta_i)|^2}{
\sigma^2[g(\theta_i)]}+2\ln\sigma[g(\theta_i)]\right],\label{log}
\end{equation}
with $\sigma[g(\theta_i)]=(1-g(\theta_i)^2)\sigma_e$.\\

\begin{figure*}
 \begin{center}
 \includegraphics[width=7cm,angle=270]{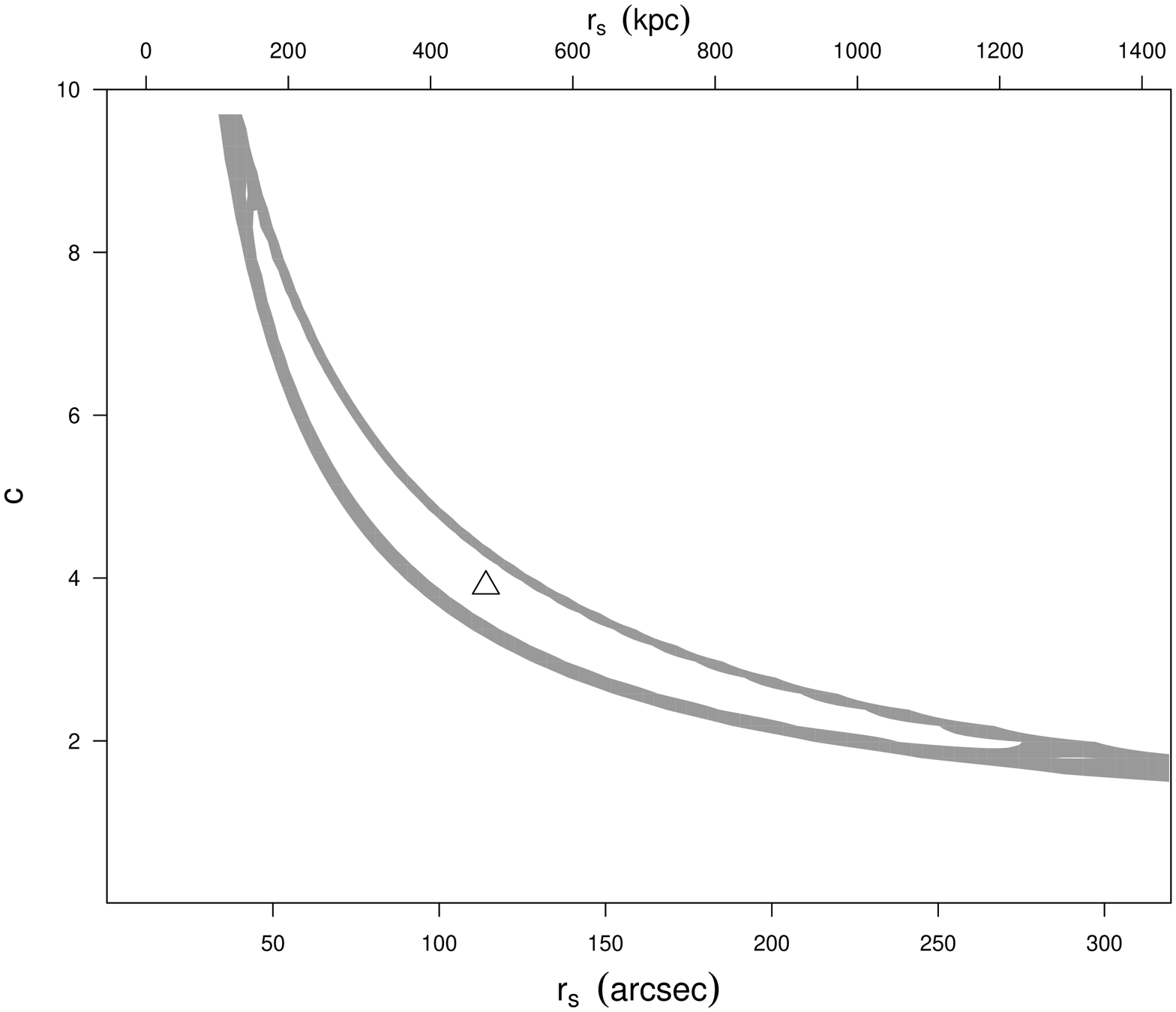}\includegraphics[width=7cm,angle=270]{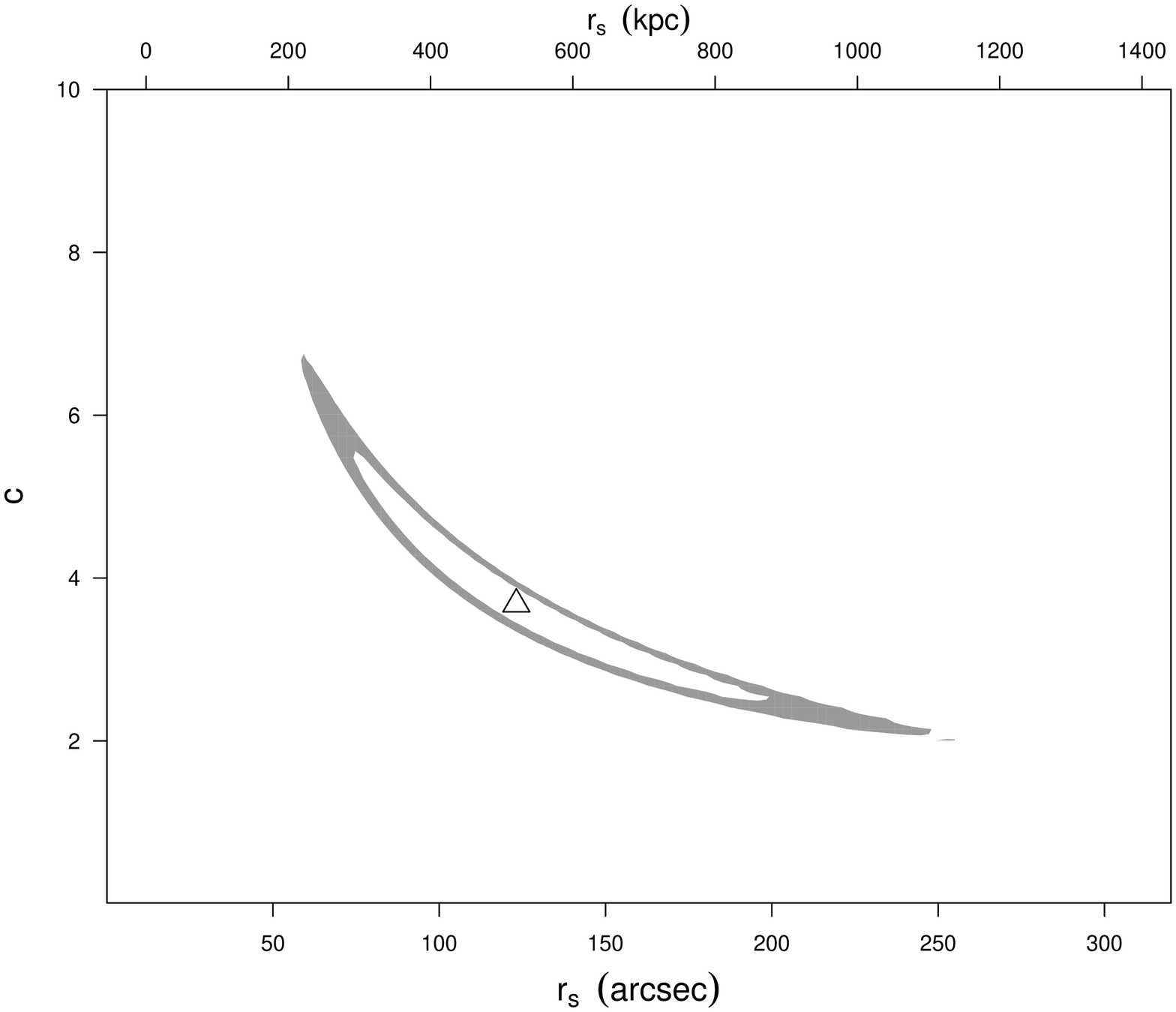}\\
 \caption{The confidence contours for NFW profile are plotted in the ($r_s$,$c$) plane, for KSB (\textit{left panel}) and Shapelets (\textit{right panel}), respectively. The confidence levels are at 68\% and 90\%, starting with the innermost one. The triangle shows the best-fit position.}\label{ll_NFW}
 \end{center}
\end{figure*}

In this analysis we assumed  both a Singular Isothermal Sphere (SIS) and a Navarro-Frenk-White (\citealt{nfw}) model.\\
In the SIS model the density profile depends on one parameter, the velocity dispersion $\sigma$: 
\begin{equation}
\rho_{SIS}(r)=\dfrac{\sigma^2}{2\pi G}\dfrac{1}{r^2}.
\end{equation}
In this profile the shear is found to be related to $\sigma$ by:
\begin{equation}
\gamma_T(\theta)=\frac{2\pi}{\theta}\frac{\sigma^2}{c^2}\frac{D_{ls}}{D_s}=\frac{\theta_E}{\theta}.
\end{equation}
(e.g. \citealt{BS01}).\\
The mass density profile predicted by the Navarro-Frenk-White model (hereafter NFW) is:
\begin{equation}
\rho_{NFW}(r)=\frac{\delta_c\rho_c}{(r/r_s)(1+r/r_s)^2}
\end{equation}
where $\rho_c=3H^2(z)/(8\pi G)$ is the critical density of the universe at the cluster redshift, $r_s$ is a scale radius related to the virial radius by means of the concentration parameter $c_{vir}=r_{vir}/r_s$ and $\delta_c$ is a
characteristic overdensity of the halo: 
\begin{equation}
\delta_c=\frac{\Delta_{vir}}{3}\frac{c^3}{\ln(1+c)-c/(1+c)}.
\end{equation}
where $\Delta_{vir}$ is the virial overdensity, approximated by $\Delta_{vir}\sim (18\pi^2+82(\Omega_{M}(z)-1)-39(\Omega_{M}(z)-1)^2)/\Omega_{M}(z)$ using the spherical collapse model \citep{BN98} for flat cosmologies.\\
The mass of the halo is:
\begin{equation}
M_{vir}=\frac{4}{3}\pi\Delta_{vir}\rho_{m} r_{vir}^3.
\end{equation}
where $\rho_{m}= \rho_{c}\cdot \Omega_{M}(z)$ is the mean density at the cluster redshift.
We solved for the mass of the cluster with the expression for the shear $\gamma_{T}(r)$ derived by \citet{bart96} and \citet{WB00} and minimized Equation \ref{log} with the \textsc{MINUIT} package.\\
Table~\ref{tab:results} shows  the best-fit parameters and the mass values derived by model fitting. The NFW profile was used keeping both the concentration and mass as free parameters (marked as NFW), and by using the relation between $c_{\rm vir}$ and $M_{\rm vir}$ proposed by \citet{bullock01} (hereafter MNFW):
$c_{vir}=\frac{K}{1+z}\left(\frac{M_{vir}}{M_*}\right)^{\alpha}$, where $M_* =
1.5\times 10^{13} /h\ M_{\odot}$, $K=9$ and $\alpha =-0.13$.\\
\begin{table*}
 \begin{center}
\begin{tabular}{llll|lll}
\hline
  &  & KSB  &   &  & Shapelets &  \\
\hline
 Parameter & MNFW &  NFW  & SIS & MNFW & NFW & SIS  \\
\hline \hline
$M_{200} (10^{14})$& $5.3_{-1.2}^{+1.4}$ & $5.6_{-2.7}^{+4.7}$ &  & $5.3_{-0.8}^{+0.8}$ & $5.9 _{-1.7}^{+2.2}$	&\\ 
$r_{200}$ (kpc)	& $1513_{-123}^{+119}$ & $1545_{-306}^{+345}$&  & $1516_{-78}^{+77}$ & $1570_{-170}^{+177}$ &	\\ 
$c_{\rm vir}$ 	& 4.51 	& $3.9_{-2.1}^{+5.6}$ &  & 4.50 & $3.7_{-1.3}^{+2.2}$ &  \\ 
$\sigma_{cl}$ (km/s)&  &  & $778_{-27}^{+26}$ &  &   &  $781_{-27}^{+26}$   \\
\hline
$red. \chi^2$  &  1.46 & 2.02 & 1.94 & 1.62 & 2.12 & 2.31 \\
   $Q$     &  0.21 & 0.11 & 0.10 & 0.17 & 0.10 & 0.06 \\
\hline
\end{tabular}
\end{center}
\caption{Mass values computed for $M_{200}$ by model fitting using a SIS profile, a NFW profile and a constrained NFW (MNFW), according with \cite{bullock01}. Best fit parameters and reduced $\chi^2$ are listed for each fit performed using both KSB and Shapelets shear catalogs, together with the goodness of fit probability $Q$. \label{tab:results} }
\end{table*} 
These values were computed assuming spherical symmetry for the cluster halo.
The effect of departures from spherical symmetry (e.g. triaxial halos) on the determination 
of the total cluster mass have been studied by several authors (e.g. \citealt{gavazzi05}, \citealt{defilippis05}).
\citet{defilippis05} showed that these effects are negligible when the mass is computed at large distances
from the cluster center, although they are important at small radii. The same authors tried to
recover a three-dimensional reconstruction of Abell 611 through a combined
analysis of X-ray and Sunyaev-Zel'dovich observations and concluded that 
the cluster was approximately spherical,  supporting our symmetry assumption.
\\
\begin{figure*}
 \begin{center}
 \includegraphics[width=7cm,angle=270]{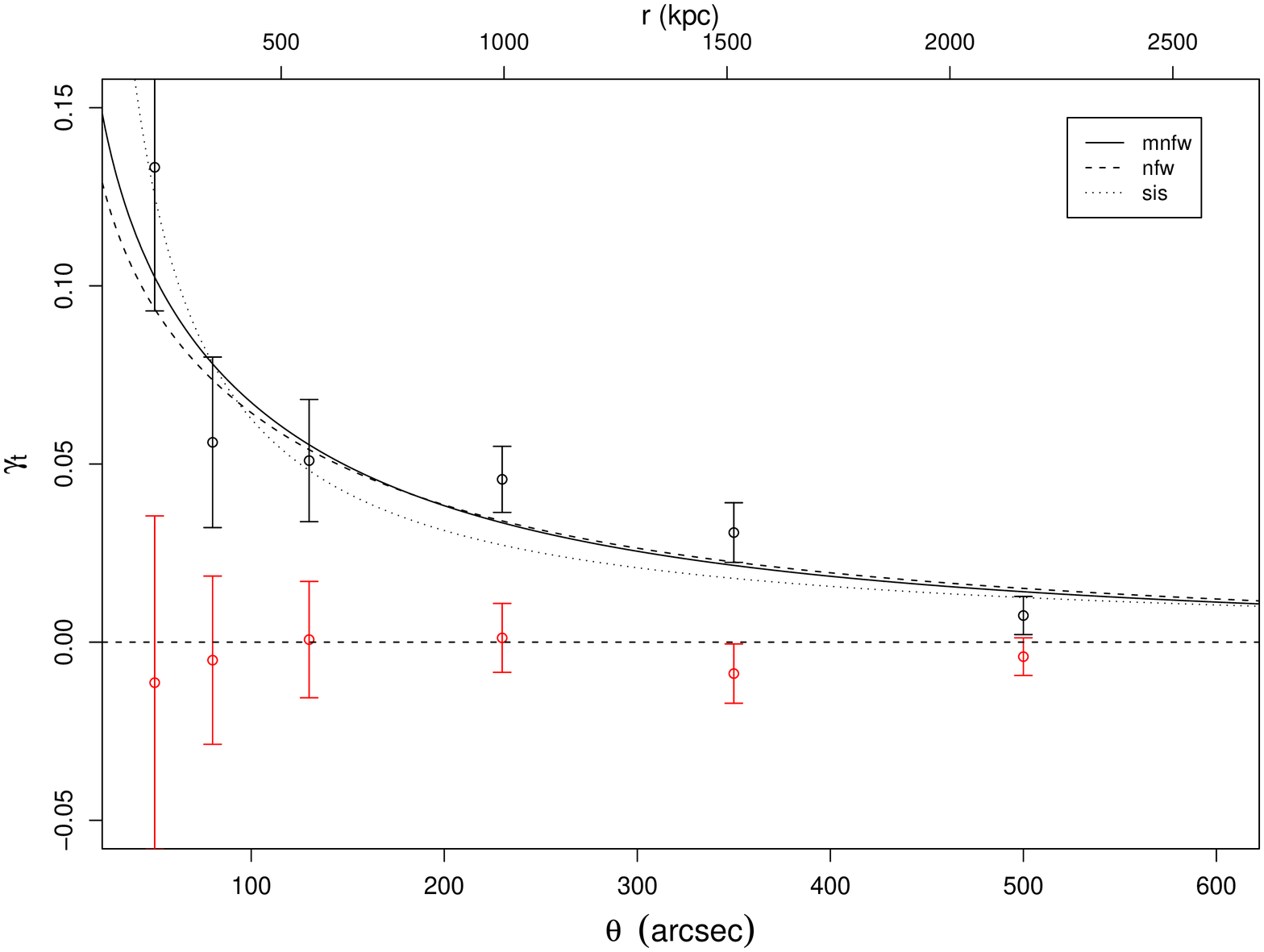}\includegraphics[width=7cm,angle=270]{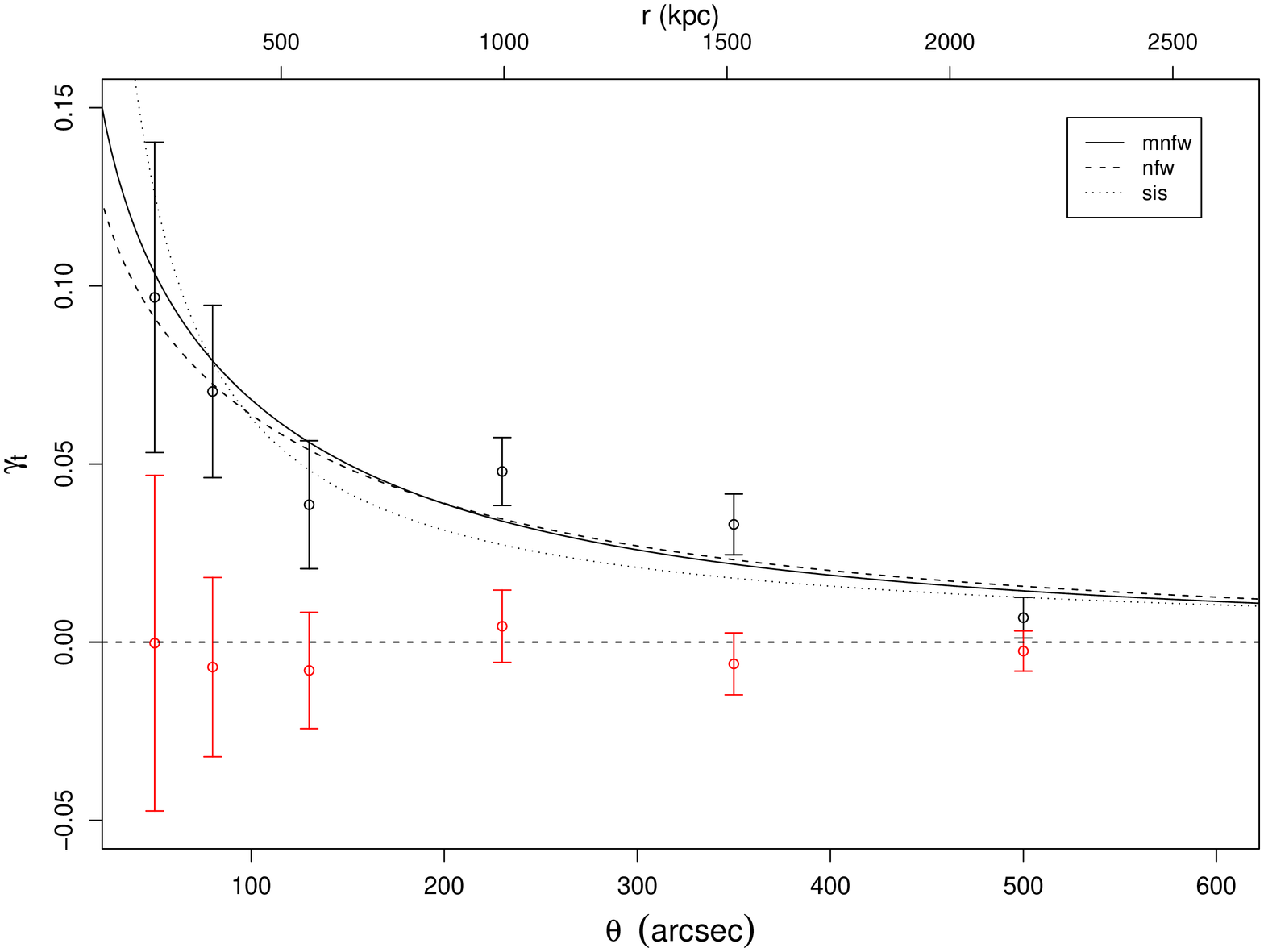}\\
\caption{Results of fitting by a NFW model ({\it dashed lines}), a constrained (M)NFW ({\it solid lines}) 
model \citep{bullock01} and a SIS profile ({\it dotted lines}), with the KSB ({\it left panel}) and Shapelets ({\it right panel}) pipeline. Average values of tangential ({\it black}) and radial ({\it red}) components of the shear, computed in logarithmic scale bins, are also plotted. Cluster masses were computed at $r_{200}$, estimated to be $\sim 350$ arcsec.}\label{fig_fit}
 \end{center}
\end{figure*}
In Figure~\ref{fig_fit} the results of fitting by a NFW model ({\it dashed lines}), a constrained (M)NFW ({\it solid lines}) model \citep{bullock01} and a SIS profile ({\it dotted lines}) are displayed, for the KSB ({\it left panels}) and Shapelets ({\it right panels}) pipeline, respectively. The black lines represent the best
fit to the unbinned data    as function of distance from the
center of the cluster. Average values of tangential ({\it black}) and radial
({\it red}) components of the shear, computed in logarithmic scale bins, are also plotted.
The latter components are expected to be zero in the absence of systematics errors.\\
 In Fig. \ref{ll_NFW} the confidence contours for the NFW profile are plotted in a plane $r_s$ vs $c$: the levels show confidence at 68\% and 90\%, starting with the innermost one.\\
The goodness of each fit ($\chi^2$) and its probability ($Q$) are also listed in Table \ref{tab:results}.

\section{Discussion}\label{summary}

We have conducted a weak lensing analysis of Abell 611 with deep $g$-band 
images from the LBC. 
Due to the complexity of the LBC PSF, we decided to use both the KSB
and Shapelets methods to measure galaxy shapes and extract the shear signal. 
KSB parametrizes the sources using their weighted quadrupole moments and is based on a
simplified hypothesis of a nearly circular PSF. In contrast, Shapelets uses a 
decomposition of the images into Gaussian-weighted Hermite polynomials and 
does not make any assumption about the best PSF model. 
Due to the large collecting area of LBT and the wide field of the LBC,  
we were able to extract a high number density of $\sim 25$ background galaxies 
per arcmin$^{2}$ over a wide field of $19\times19$ arcmin$^2$ and this 
allowed us to perform an accurate weak lensing analysis.\\
The two shear catalogs, derived by the KSB and Shapelets pipelines, were matched and 
common sources (with a number density of $\sim$ 23 galaxies/arcmin$^2$) were used to estimate mass measurements for Abell 611 by two different 
weak lensing techniques: aperture densitometry and a parametric model fitting. 
In both approaches we assumed that the BCG was the center of the mass
distribution, which is supported by S-Maps (see Fig. \ref{fig:snr_ksb}, Section~\ref{s-map}).\\ 
The projected mass values obtained by aperture densitometry within a radius of $\sim1500$ kpc are: 
$7.7\pm{3.3}\times 10^{14} M_{\odot}$ and $8.4\pm{3.8}\times 10^{14} M_{\odot}$, using KSB 
and Shapelets, respectively. These estimations are model independent \citep{clowe98}, but they are affected 
by large uncertainties. As discussed in Section \ref{apdens}, the contribution 
of uncorrelated structures along the line of sight can be source of noise for aperture measurements, although they can be decreased by averaging the results for several clusters (\citealt{hoekstra01}) or 
corrected by using photometric redshifts of the sources, if available. \\
Table~\ref{tab:results} shows the results of fitting the observed shear 
with a parametric model. We assumed both a SIS and a NFW mass density profile. 
We first fitted a NFW model leaving both $c$ and $r_s$
as free parameters: as displayed in Fig. \ref{ll_NFW}, the uncertainty on these
two parameters provided by the fit is high and does not allow to put a
strong constrain on the concentration. Nevertheless, the best-fit value
($c \sim 4$)  is in good agreement with the one obtained when the Bullock
et al. (2001) relation is adopted ($c = 4.5$).
For the NFW fits the quoted masses are $M_{200}$, within the radius $r_{200}$ where the density is 200 times the critical density.
The estimated value of $r_{200}$ for Abell 611 is $\sim1.5$ Mpc ($5.8'$).\\
   The weak lensing mass measurements obtained from both the KSB and Shapelets shear catalogs are in agreement, within uncertainties: by using a (M)NFW profile we obtain
$M_{200} \sim 5.3_{-1.2}^{+1.4} \times 10^{14} M_\odot$ and $5.3_{-0.8}^{+0.8} \times 10^{14} M_\odot$, respectively.
The smaller uncertainties of Shapelets results show that this method provide a higher accuracy than KSB.
Moreover, the goodness of the fits ($\chi^2$) in Table~\ref{tab:results} shows that both Shapelets and KSB provide a best-fit with higher probablity ($Q$) by using a NFW mass density profile than a SIS profile.  \\
Abell 611 was previously targeted for a weak lensing study by \citet{dahle06},
who used $V$ and $I$ observations from several facilities to target a 
large number clusters (see \citealt{dahle02} for more details). 
They used the KSB \citep{KSB95} shear estimator as described in \citet{kaiser00} to derive 
the shear signal from the images. The mass of the cluster was then derived by fitting the 
observed shear with a NFW mass density profile \citep{nfw}. They
assumed a concentration parameter as predicted by \citet{bullock01} and obtained
a mass value of $M_{180} = (5.21\pm 3.47)\times10^{14}h^{-1}M_{\odot}$  within
$r_{180}$, the radius within which the density is 180 times the critical density.
A more recent mass estimate of Abell 611 is presented in \citet{PD07}, who used
the data collected in \citet{dahle06} and obtained a value of $M_{500} = 3.83\pm
2.89\times10^{14}h^{-1}M_{\odot}$  within $r_{500}$, the radius within which the 
density is 500 times the critical density. In these papers the authors 
extrapolated the NFW profile out to $r_{500}$ because the data were 
insufficient to extend this far in projection from the cluster center. 
For their work on Abell 611 $r_{fit}/r_{500}=0.59$ (note
$r_{500}=0.66\times r_{200}$). \\ 

Our weak lensing estimates for the mass of Abell 611 are in agreement with the
previous results of \cite{dahle06} and \cite{PD07}, but the depth and the larger area
covered by LBC data allowed us to perform a more accurate analysis.\\
A recent weak lensing analysis of Abell 611 has recently been done by \cite{okabe09}
using Subaru/Suprime-Cam observations in two filters ({\it i'} and {\it V}).  The authors used the color ($V-i'$)
information to select the background galaxies to use for their cluster lensing analysis, getting a galaxy number density of 
$\sim 21$ galaxies/arcmin$^2$, with a SNR $\sim 10$. By fitting the 
mass density profile of the cluster using a NFW model, they found a mass value 
$M_{200}= 5.13^{+1.16}_{-1.00}\times 10^{14}M_\odot$ and a $c_{vir}=4.14^{+1.73}_{-1.21}$ 
(private communication), which are in good agreement with our results.\\
   A new mass estimation of Abell 611 was performed by \cite{newman09} over wide range of cluster-centric distance (from $\sim$ 3 kpc to 3.25 Mpc) by combining weak, strong and kinematic analysis of the cluster, based on Subaru, HST and Keck data, respectively. They found a mass value $M_{200}= 6.2^{+0.7}_{-0.5}\times 10^{14}M_\odot$ with $c=6.95\pm{0.41}$ by using a NFW model fitting, in agreement with our results. We note that such a large value of $c$ cannot be rejected by our NFW fits, due to the large uncertainties on NFW parameters (see Tab.\ref{tab:results} and Fig.\ref{ll_NFW}). \\ 
Finally, we also compared the mass values obtained by our weak lensing analysis to X-ray estimates of $M_{200}$ available in literature.
\cite{SA07} analyzed Chandra data of several clusters and modelled their total mass profile (dark plus luminous matter)
using a NFW profile. They found for Abell 611 a scale radius $r_s = 0.32_{-0.20}^{+0.10}$ Mpc and a 
concentration parameter $c = 5.39_{-1.51}^{+1.60}$, which provide a mass value
$M_{200} \sim 8\times 10^{14}M_\odot$ at $r_{200}\sim 1.7$ Mpc, in agreement, within the uncertainties, with the results obtained by us.
A more recent X-ray analysis of Chandra observations of Abell 611 has been
performed by Donnarumma et al. (in preparation). They obtained
a value of $M_{200}= 1.11\pm 0.21\times 10^{15}M_\odot$ at $r_{200}\sim
1900$ kpc, with $r_s = 407_{-86}^{+120}$ kpc and $c = 4.76_{-0.78}^{+0.87}$.
Their projected mass at $r_{200}$ is $1.28 \pm 0.24 \times 10^{15}M_\odot$. Such
value is in agreement, within the statistical uncertainties, with the mass measured by
aperture densitometry, but higher than the value estimated by the parametric model. Additional information 
on the mass will be derived from a strong lensing analysis of Abell 611 (Donnarumma et al. in preparation). \\
 This work shows that LBT is a powerful instrument for weak lensing studies, but we want to stress that the data here analyzed did not allow us to use the full capabilities of the telescope. The presence of bright saturated stars hampered to use the whole field of the camera for the analysis of Abell 611. 
In addition, the analysis of weak lensing is expected to be improved by
the usage of the Red Channel
in LBC, which was not yet available during these observations. The
present results are nevertheless important
to demonstrate the capabilities of LBC for weak lensing; we therefore
plan to extend to other clusters such analysis,
now using the Red Channel.\\

\begin{acknowledgements}
Observations were obtained with the Large Binocular Telescope
at Mt. Graham, Arizona, under the Commissioning and Science Demonstration
phase of the Blue Channel of the Large Binocular Camera. The LBT is an international
collaboration among institutions in the United States, Italy and Germany.
LBT Corporation partners are: The University of Arizona on behalf of the Arizona
university system; Istituto Nazionale di Astrofisica, Italy; LBT
Beteiligungsgesellschaft, Germany, representing the Max-Planck
Society, the Astrophysical Institute Potsdam, and Heidelberg
University; The Ohio State University, and The Research Corporation,
on behalf of The University of Notre Dame, University of Minnesota and
University of Virginia.\\
This paper makes use of photometric redshifts produced jointly by Terapix and the VVDS team. \\
Part of the data analysis in this paper was done using the R software (http://www.R-project.org).\\
We thank the anonymous referee for useful comments and suggestions which improved 
the presentation of this work.
AR acknowledges the financial support from contract ASI-COFIS  I/016/07/0.
LF, KK and MR acknowledge the support of the European Commission Programme 6-th framework, Marie Curie Training and Research Network “DUEL”, contract number MRTN-CT-2006-036133. LF is partly supported by the Chinese National Science Foundation Nos. 10878003 \& 10778725, 973 Program No. 2007CB 815402, Shanghai Science Foundations and Leading Academic Discipline Project of Shanghai Normal University (DZL805).
AD, SE, LM, MM acknowledge the financial contribution from contracts ASI-INAF I/023/05/0 and I/088/06/0.
SPH is supported by the P2I program, contract number 102759.\\
\end{acknowledgements}

\bibliographystyle{aa}

\end{document}